# Analytics and Machine Learning Powered Wireless Network Optimization and Planning

Ying Li, Djordje Tujkovic, Po-Han Huang
Meta Platforms, Inc.

## Abstract

It is important that the wireless network is well managed, optimized and planned, using the limited wireless spectrum resources, to serve the explosively growing traffic and diverse applications needs of end users. Considering the challenges of dynamics and complexity of the wireless systems, and the scale of the networks, it is desirable to have solutions to automatically monitor, analyze, optimize, and plan the network, instead of the traditional way of engineers manually monitoring, analyzing, tuning, and planning the network.

This article addresses the limitations of existing network optimization and planning technologies by providing approaches and solutions of data analytics and machine learning (ML) powered optimization and planning. The approaches include analyzing some important metrics of performances and experiences, at the lower layers and upper layers of open systems interconnection (OSI) model. The approaches include deriving a metric of the end user perceived network congestion indicator. The approaches also include monitoring and diagnosis such as anomaly detection of the metrics, root cause analysis for poor performances and experiences. The approaches also include enabling network optimization with tuning recommendations, directly targeting to optimize the end users experiences, via sensitivity modeling and analysis of the upper layer metrics of the end users experiences v.s. the improvement of the lower layers metrics due to tuning the hardware configurations. The approaches also include deriving predictive metrics for network planning, and modeling of traffic demand distributions and trends, incentives of traffic gains if the network is upgraded, etc. The models detect and predict the suppressed engagement or suppressed traffic demand.

These approaches of optimization and planning may provide more accurate detection of optimization and upgrading/planning opportunities for cells at a large scale, enable more effective optimization/planning of networks, such as tuning cells configurations, upgrading cells' capacity with more advanced technologies or new hardware, adding more cells, etc., improving the network performances and providing better experiences to end users of the networks.

## Introduction

Wireless networks are serving explosively growing traffic of end users [1]. To support the growing traffic, the wireless networks are getting more and more advanced and complex, as well as more wireless network access points are being deployed, such as cellular cells, WiFi access points, etc..

In wireless networks, one of the key challenges is that wireless systems have to be very well managed, optimized and planned, especially for the reliability/latency/rate, as well as the interferences, otherwise, the performance might not be up to the level that could well tradeoff the cost. Wireless connections could potentially be more dynamic than the wireline connections, due to the wireless environment.

Considering the challenges of dynamics and complexity of the wireless links and systems, as well as the potential large scale of the network, it is desirable to have automated solutions to automatically monitor, analyze, optimize, and plan the network, instead of the traditional way of engineers manually monitoring, analyzing, tuning, and planning the network.

The automated optimization consists of the chain of the data collection, monitoring, storage, training, modeling, tuning, planning, etc., to automatically optimize and plan the network performances and end users' experiences, to minimize the manual involvement. This can not only boost the performance of the network, but also further save the operation/maintenance expenses (OPEX), as well as potential capital expenditures (CAPEX) saving (the optimization could also be done by minimizing needed wireless access nodes, or needed advanced hardware, to serve the areas).

Mobile network operators (MNOs) may monitor network traffic of a large number cellular towers at network infrastructure level, optimize and plan network performance based on the monitored network traffic. However, the MMOs might lack means to detect whether users are happy in network experiences, for those users supported by the large number of cells of cellular networks, hence the optimization or planning might not be easy to directly target for better users' experiences.

Traditional methods of measuring network performance by driving a measurement vehicle around in the network is very inefficient and limited in capability. Furthermore, the optimization and planning that only rely on network infrastructure level information may have many limitations and could lead to suboptimal results.

For example, although the MNOs may monitor network and resource utilization (e.g., throughput, traffic volume, radio resource utilization percentage, etc.) at network infrastructure level or lower layers of open systems interconnection (OSI) model, this infrastructure level information may not fully reflect the actual user experience and application usages at application level. The MNOs might lack means to directly evaluate the end users experiences, Quality of Experiences (QoE) of end users because of a lack of effective means to access application usage data at application level. The MNOs might have challenges of associating network lower layer performance degradation with users' experiences at higher layers such as application layer.

When lower layers and upper layers are analyzed separately, the optimization and planning may have suboptimal results with limited effectiveness and impact. For instance, sometimes even if the lower layers are improved by a large extent, however, it might not always be translated to better end users' experiences at application layer, if the users' experiences are bottlenecked by other factors than the lower layers.

This article discusses approaches and solutions which provide data analytics and machine learning (ML) powered optimization and planning for wireless networks. These approaches address the limitations of existing network optimization and planning technologies by providing analytics and ML powered technical approaches and solutions for monitoring and diagnosis such as anomaly detection of the metrics, root cause analysis, and optimization with tuning recommendations, as well as planning with predictive metrics.

The approaches include collecting deidentified application data (anonymized data, i.e, users cannot be identified from the data) and obtaining important metrics of performances and experiences, such as users quality of experience, download speed, latency, etc. at upper layers of OSI model, and radio signal strength, signal quality, etc. at the lower layers. The approaches also include deriving a metric of the end user perceived network congestion, and having this congestion metric as part of the important metrics.

The approaches include anomaly detection (AD) of the important metrics in one or more areas covered by a network and sending alerts to the MNOs of the network as the basis for network optimization and planning. The approaches include root causes analysis (RCA) of the poor or degraded network performances and user experiences. The approaches also include enabling network optimization with tuning recommendations, based on information (e.g., infrastructure level, application level) from multiple layers of OSI model, as well as location information, instead of information from one single layer only, and therefore enable better network optimization and planning.

The approaches include deriving predictive metrics for network planning, where the traffic demand distributions and trends, end users experiences, incentives of traffic gains if the network is upgraded, etc., are analyzed and modeled. The models detect and predict the suppressed engagement or suppressed traffic demand, possibly indicating the insufficient capacity hence the need for densification and upgrading. The planning directly targets for better users' experiences, hence this is end users' experiences driven planning, with the metrics measured at the upper layers in the picture.

These approaches of optimization and planning may provide more effective recommendations with reduced computational cost for improving the network performances and end users experiences. The approaches may provide faster and more accurate detection of optimization and upgrading/planning opportunities for cells at a large scale, enable more effective optimization/planning of networks (e.g., tuning cellular towers, upgrading cells' capacity with more advanced technologies or new hardware, adding more cells, etc.), and providing better experiences to end users of the networks.

The remainder of the article discusses details regarding the ML powered optimization and planning for the wireless networks.

# System Descriptions

In this section, we provide some descriptions of an example of the system, where the Machine Learning (ML) powered approaches for network optimization and planning can reside.

## System overview

Figure 1 shows an example of the Machine Learning (ML) powered wireless networks, with the modules of data collection and processing, ML powered network optimization and network planning.

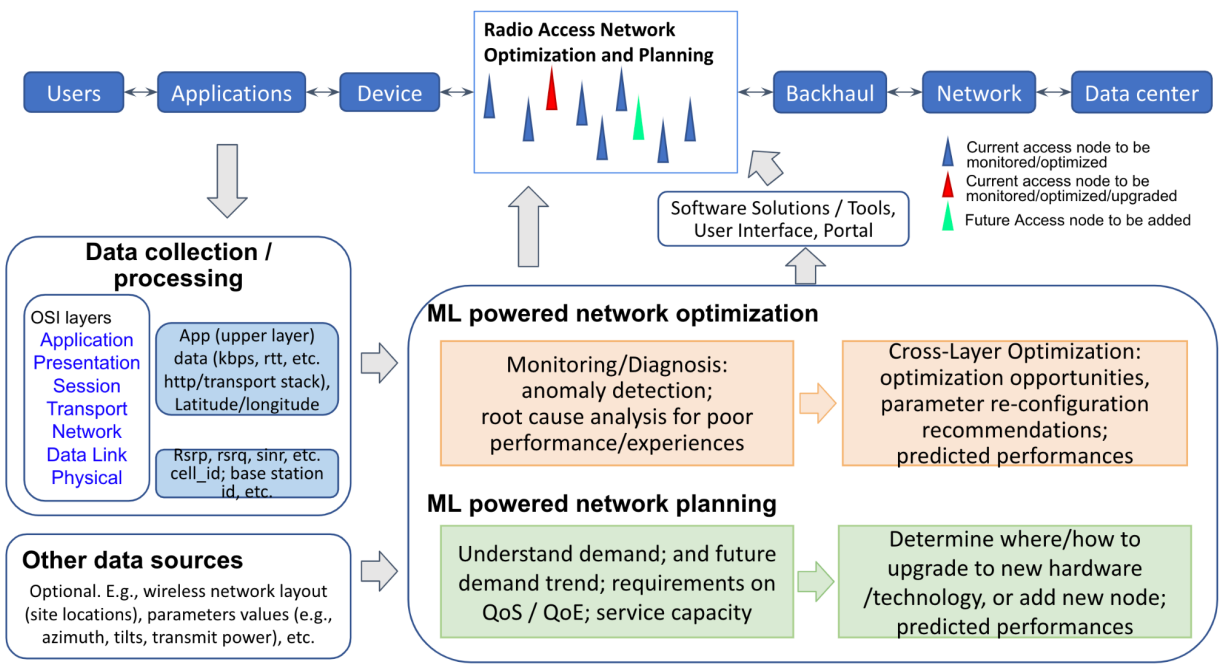

*Fig. 1 An example of the Machine Learning (ML) powered wireless networks, with the modules of data collection and processing, which is the input for ML powered network optimization and network planning, whose output can be used for Radio Access Network (RAN) optimization and planning.*

In the figure, a user uses user equipment (UE, the device). Application (app) software can be run on the device. The UE is connected to the Radio Access Network (RAN), which is connected to the core network via backhauls. The backhaul and core network could be owned by MNO. The MNO core network then can be connected to the Internet, or connected to edge networks with some content caching functions first then go to the Internet. The content that a user wants to access via the app could be stored in the edge network or data center. The data center could be owned by the enterprise who owns the app. If a user requests some content and the content is at the edge of the network, then the content can be delivered to the user via the edge, core, backhaul, RAN, and to the device. If the content is at the data center, then, the content request would go beyond the edge network, and go deeper towards the servers in the data center, then the data center can send the content through the network, MNO core and backhaul, RAN, and then to the device.

The system may include the modules of data collection and processing, and other data sources optionally. The data can be the inputs for the ML powered network optimization and planning, which output decisions or recommendations of the optimization and planning, feeding to RAN. The network optimization and planning may be also as solutions and tools, via User Interface (UI) or portal for visualization, being provided to the customers such as MNOs. An MNO can monitor and diagnose the network performances and end-users experiences such as at daily basis, see the solutions, recommendations of the tuning, time series, maps, predicted performances, and the MNO can follow up to take actions to tune, and keep on monitoring, optimize and plan the network, also possibly interacting with the UI or portal to input the actions having been taken, time stamps of the actions taken. The system forms a closed loop of the solutions, and it can iterate.

The deidentified data can be collected and stored in an entity on the Internet (where the entity could be a server, a part of a cloud, etc.). The data collection can be done, via the application (app). Over the open systems interconnection (OSI) model, the data collected could be at app (upper layer), which can be used to derive the download speed (kbps), round trip time (rtt), with the HTTP (app) and the Transport stack. The location data might be also collected if applicable. The data collected could be at the lower layer, such as reference signal received power (rsrp), reference signal received quality (rsrq), the cell identifier of the cell which is serving this UE, the base station identifier of the cell.

The data then can be processed, such as cleaning the data, aggregating the data over a time window of hours, daily, days, etc. Then, the data can be used as the input to the Machine Learning (ML) powered network optimization and planning, which can be used to provide optimization and planning for RAN.

The other data sources may include wireless network layout (site locations), parameters values (e.g., azimuth angle, tilts, transmit power), etc. Some of the other data sources may be optional.

The ML powered network optimization modules and planning modules can include data analytics, modeling, machine learning, wireless domain knowledge-based rules, etc. The rules based approaches and the ML based approaches can be combined.

The ML powered network optimization modules may include monitoring and diagnosis for poor network performances or experiences, with anomaly detection (AD), root cause analysis (RCA) for poor performance/experiences, and so on. They may also include cross-layer optimization, with the sensitivity modeling of the lower layer metrics (such as radio signal strength or quality, etc.) v.s. the upper layer metrics (such as app layer user experienced download speed, latency, etc.), where the cells with higher sensitivity of the lower layers to higher layers can be for higher priority as the optimization opportunities of tuning the lower layers metrics. They may also include optimization tuning modules, where some of the main parameters tuning and re-configuration recommendations are made, and predicted performances if tuning are computed and provided. The main parameters may include the azimuth angle, tilts, transmit power, etc.

The ML powered network planning modules may include demand modeling, which is to understand the demand, and future demand trends. They may also include the understanding of the requirements on quality of services (QoS) and quality of experiences (QoE), such as the requirements of application level of the download speed, user's end-to-end round trip time, etc. The planning modules may also include the modeling of the current capacity the network can provide, and detect whether the current capacity of the network is no longer sufficient to the demand, hence the end-users demand has been suppressed. The planning modules may also include the predictive modeling on determining where/how to upgrade to new hardware /technology, or add new nodes, with predicted performances.

## Optimization levels

Figure 2 illustrates the optimization levels. Each level can be corresponding to different granularity, and time scale.

The data analytics will also consider the granularity and time scale. Especially we will pay attention to the assumptions of which configurations are assumed to be fixed/settled when tuning other configurations. This is a way of decoupling. However, joint optimization could be also an option, if practically the time scale is appropriate.

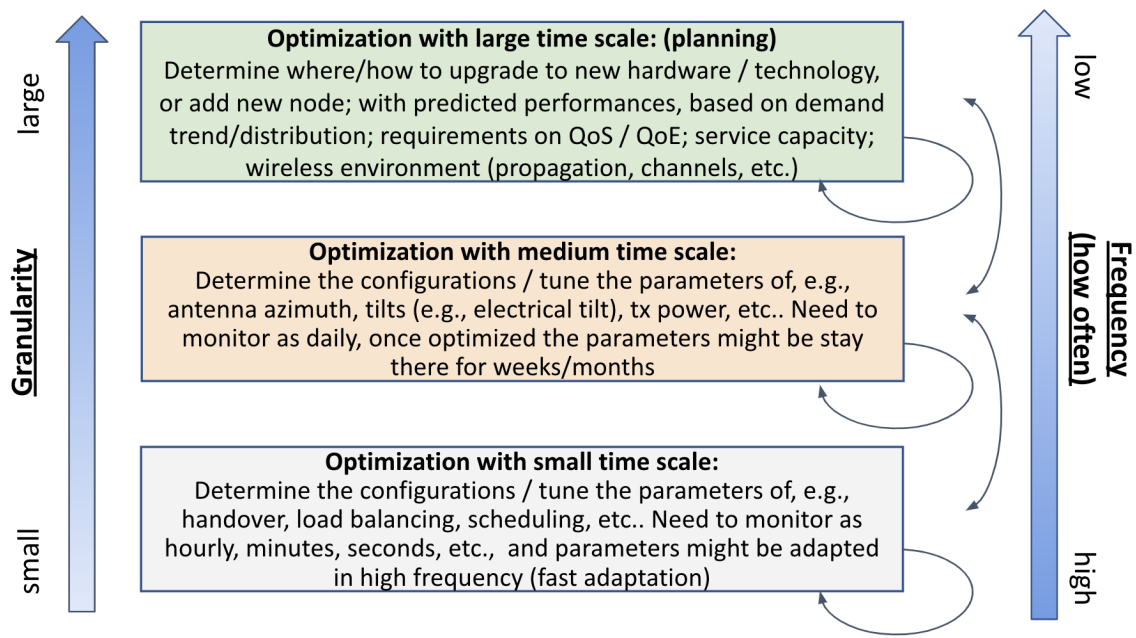

*Fig. 2 An example of the illustration of wireless network optimization levels.*

In the figure, as an example, three levels of the optimization granularity and time scales (i.e., how often or how frequently the optimization is performed) are illustrated.

Network planning can be viewed as an optimization with a large time scale, and a large granularity in geographic area and actions to take. The planning can determine where/how to upgrade to new hardware or new technology, or add new nodes, with predicted performances. The decisions can be based on modeling and analysis on the demand trend/distribution, requirements on QoS / QoE, service capacity, wireless environment (propagation, channels, etc.) and so on.

An optimization with a medium time scale could be to determine the configurations and tune the parameters of, e.g., antenna azimuth, tilts (e.g., electrical tilt), transmit power, etc.. It may need to be monitored daily/hourly. Once optimized the parameters might stay there for weeks or months.

An optimization with a small time scale could be to determine the configurations and tune the parameters related to, e.g., handover, load balancing, scheduling, etc.. It may need to be monitored hourly, minutes, seconds, etc., and parameters might be adapted more frequently (fast adaptation).

Each level could run iterations on its level for optimization. Higher frequent optimization could assume the lower frequent one is settled (for decoupling), otherwise joint optimization might be considered.

# ML-Powered Wireless Network Optimization

ML-powered network optimization modules as in Fig. 1 plays an important role for wireless networks performance improvement and end users experiences improvement. This section provides more details regarding network optimization, mostly at the medium time-scale level of the optimization as in Fig.. 2, where it could provide a base or foundation for further small time-scale level of the optimization, as well as a pre-step before going to the possibly more expensive solutions as in the network planning (the large time-scale level of the optimization).

## Metrics of performances and experiences indicators

At the upper layers, some of the metrics of performances and experiences indicators may include the following.
- The download speed (kbps) derived at the application layer of the HTTP stack, which can be measured as the speed per http request response, calculated as the total bits of the http request response, dividing the time duration of time-to-first-bit to time-to-last-bit.
- The latency, measured as the round trip time (rtt), with the HTTP (app) and the Transport stack.
- The congestion indicator that an end user might perceive, maybe derived from the download speed user experiences at app layer, across busy hours and non-busy hours. (More details are provided in the next subsection).
- The video playback metrics, measured at app layer, such as video stall ratio, the stall time before the playback starts, playback success rate, mean time in-between rebuffering, mean opinion score, etc.
- Note that these metrics are measured at the upper layers, hence all the possible problems at the layers below the layer where these metrics are measured, might be captured or reflected in these metrics.
- Also note that the download speed measured at app layer, is not the throughput measured at the lower layers, such as at physical (PHY) layer. The download speed at app layer could capture or reflect all the layers below app layer, including the throughput and the delay at PHY or MAC layer, the transport layer throughput and delay, the content delivery network throughput and delay, and all the buffers along the information delivery path.

At the lower layers, some of the metrics of performances and experiences indicators may include the following [2, 3].
- The wireless radio signal strength. This can be the reference signal received power (RSRP). In 4G LTE it is RSRP, and in 5G it includes synchronization signal RSRP (SS RSRP), channel state information RSRP (CSI RSRP). This category of the metrics can be very related to the cell coverage.
- The wireless radio signal quality. This can be the reference signal received quality (RSRQ). In 4G LTE it is RSRQ, and in 5G it includes synchronization signal RSRQ (SS RSRQ), channel state information RSRQ (CSI RSRQ). This category of the metrics can be very related to the cell capacity.
- The wireless radio signal-to-interference-and-noise-ratio (SINR). In 4G LTE it is SINR, and in 5G it includes synchronization signal SINR (SS SINR), channel state information SINR (CSI SINR). This category of the metrics can be very related to the cell capacity.

## Congestion indicator based on user experiences (app layer data)

A congestion indicator metric related to how a user might be perceiving congestion in the network can be derived, by looking into the upper layer metrics, at busy hours and non-busy hours.

For the app layer download speed (which is closer to users experiences), a user might experience lower download speed at app layer at busy hours v.s. non-busy hours. So, to capture the perspective of app layer download speed, it is not only about a metric of the download speed, but also about a metric of the busy hour download speed, and in addition, a metric to measure the degradation of download speed at busy hours v.s. non-busy hours.

The congestion indicator at app layer captures the degradation of download speed at busy hours v.s. non-busy hours. It is calculated as the following, for each cell.

$$\text{Congestion indicator} = (A - B) / A * 100\%$$
- If $A > B$, $0 < \text{congestion indicator} < 100\%$
- If $A < B$, then congestion indicator $< 0$; enforced to 0
A = download speed at non-busy hours
B = download speed at busy hours

where busy hours/ non-busy hours are derived based on the aggregated hourly number of samples, for a respective cell.

The naming of "congestion indicator" here, is motivated by how an end user would be detecting or perceiving "congestion", or how an engineer would infer user's experiences of "congestion", by using the app layer data. A question is: When might a user be saying, "I experience congestion, and I know it is the network problem"? If a user experiences really good download speed at some hours and then the user experiences really bad download speed at some other hours, and the user might be unhappy, and the user might say, this could be "the network is congested" at these hours that the download speed is really poor, as the download speed is good at some hours (the device might be ok). An ML model can be used, to derive the threshold of this congestion indicator, larger than which the congestion at the app layer might be noticeable by end users or engineers.

When the congestion indicator is high, the user might be unhappy about this mobile operator; the user might complain to the operator, and hope this operator would upgrade the network. In some of our studies, this congestion indicator derived at app layer, as well as the download speed at app layer, have a relationship with the user's churn away from the mobile operator. The lower the download speed is, and the higher the congestion indicator is, the higher likelihood of the churn might be.

Note that the congestion indicator metric derived via app layer data, is not to replace or compete the congestion metric derived at the lower layers (e.g. traffic load, physical radio resource usage percentage, etc.), rather, this is an app layer metric, and could be related to user's perception and experiences, users' happiness about experiences, and this could be complementary to the congestion derived at the lower layers.

The congestion indicator can be derived at the cell level, as each cell might have its own service capacity. The cell level congestion indicator then can be extend it from cell to geographic areas, for example, a weighted averaged congestion indicator can be derived for an area, where the congestion indicators of the cells overlapping with the area can be averaged using weights, where the weights can be the number of samples of data in each cell.

## Anomaly Detection (AD) and Root Cause Analysis (RCA)

Anomaly detection (AD), root cause analysis (RCA), and tuning recommendations are among the beneficial functions for network optimization, such that the network resource utilization can be efficient, while end users' experiences can be guaranteed and improved. The trend is to use data-driven approaches, to replace the traditional manual process. In [4], an example of ML assisted AD and RCA is provided.

Traditionally, these AD, RCA, tuning recommendation functions could be done manually by the engineers, which might involve a lot of engineers' time, engineers eye-balling the excel-sheets, and field engineers driving tests, etc. The traditional approaches are computationally costly, not scale well, not very easy to find the optimal solution either, especially when the wireless network becomes more and more complicated with more and more tuning knobs, these drawbacks are more prominent.

The bright sides are that data-driven approaches, automation, and machine learning are booming, which gives or will give the Telecom industry the right disruptive paradigm shifts. Data-driven approaches are of great value to do automation (not the traditional manual approaches) in the Telecom industry.

Anomaly detection (AD) can monitor the network (or the health of the network), and detect anomaly (or 'symptom'), where the anomaly can include, e.g., the anomaly of end user experience metrics (e.g. QoE, end-user's download speed, latency, congestion indicator, etc, measured at upper layers), the anomaly of the network performance metrics (e.g. the radio signal strength, signal quality, etc.).

The anomaly can be in the spatial dimension (e.g., at certain spot or in certain area, or in certain cell), or it can be in the time dimension (e.g., at which time period), or combined spatial and temporal (e.g., cell-time (which cell(s) and which time window(s))). The detected anomaly can be for poor metrics, or for metrics degradations. These anomalies can be sent or notified to the relevant parties, such as MNOs.

Root cause analysis (RCA) can further dig deeper to find (or 'diagnosis') the causes or reasons underneath the 'symptoms' (e.g., the anomaly that AD detected), i.e., finding the major or critical causes which results in the anomaly. For instance, for certain download speed being low of a cell, the root causes may include, e.g., the signal strength being low (hence the cause being coverage), the interference being high (hence the cause being interference), the cell is too congested (hence the cause being capacity), etc.

The RCA can identify the top causes, and it can also (if possible) rank the causes according to its contribution to cause the anomaly. By ranking the causes, it can be helpful for the 'treatment' in the next step, i.e., optimizing the network. In other words, the RCA finds the 'knobs to tune' to optimize the network, and by ranking the causes, it finds out the most effective knobs to be tuned, so the network can be optimized (i.e., the anomaly can be 'cured').

For RCA, wireless domain knowledge based rules can be used, as a simpler version of the root causes identification. More advanced root cause analysis can be done, e.g., via the sensitivity analysis, with machine learning regressions. Some of the details of sensitivity analysis are as in the next subsection.

Figure 3 illustrates an example of network optimization of the anomaly detection, root cause analysis, and sensitivity analysis.

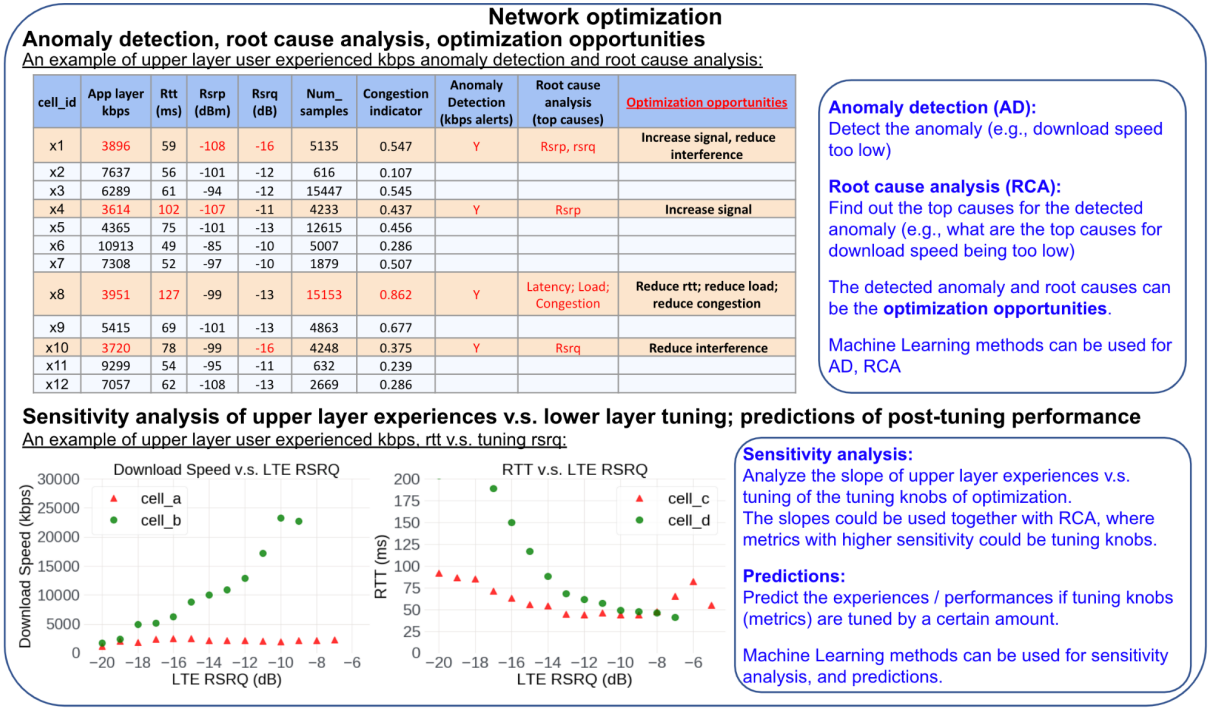

*Fig. 3 An example of network optimization of the anomaly detection, root cause analysis, and sensitivity analysis. Cell_id means identifier of a cell. App layer kbps means the download speed in the unit of kilobits per second (kbps) measured at app layer. Rtt in the unit of millisecond (ms) means the round trip time the user experienced at upper layers. Rsrp, rsrq is the reference signal received power (unit dbm), reference signal received quality (unit dB), respectively. All these metrics are of the median values of the samples measured at a cell in a time window such as several days. Num_samples means the number of data samples measured at this cell. Congestion indicator is (download speed at non-busy hours - download speed at busy hours) / download speed at non-busy hours.*

In Fig. 3, for AD and RCA, an example of the upper layer user experienced download speed low as the anomaly is provided. In the example, the cells with lowest app layer download speed are identified and highlighted as the cells with download speed low anomaly. In the figure, Cell_id means identifier of a cell. App layer kbps means the download speed in the unit of kilobits per second (kbps) measured at app layer. Rtt in the unit of millisecond (ms) means the round trip time the user experienced at upper layers. Rsrp, rsrq is the reference signal received power (unit dbm), reference signal received quality (unit dB), respectively. All these metrics are of the median values of the samples measured at a cell in a time window such as several days. Num_samples means the number of data samples measured at this cell. Congestion indicator is (download speed at non-busy hours - download speed at busy hours) / download speed at non-busy hours.

There are four cells in the example in Fig. 3, identified as with the lowest download speed. The anomaly could be determined by using some rules, such as the cells with the lowest M% among all the cells in the scope (the scope could be, e.g., a geographic area, with similar frequency bandwidth, etc.), e.g., M% being 5%, or 10% etc., or download speed below a preset threshold, like 4 Mbps in this example.

The RCA in the example, gives the potential causes. To determine the potential causes, in this example, the anomaly of the metrics other than download speed is also detected, e.g., with the worst N% of the cells, e.g., N >=M, and N% being 10%, 20%, etc. The rationale is that for a metric Y being bad, among other metrics X1, X2, X3, …, if X2 is bad but X1 and X3 are good, then the possible cause might be X2. The reason for N being no smaller than M is to provide some margin or room for having more potential causes to be the candidates, not too restrictive so as to pre-exclude the actual causes.

Once anomaly is identified, and the potential causes are identified, the potential tuning or optimization opportunities can be also identified.

# Optimization towards users experiences, via sensitivity of experiences metrics v.s. lower layer metrics

The sensitivity relationships analysis of app layer metrics v.s. lower layer metrics enables users-experiences-driven ML-powered network optimization, where the optimization approaches can optimize networks directly towards better users' experiences.

This means that users' experiences are one of the important targets out of the ML-powered network optimization.

As the data collection and processing described in earlier sections, the data include metrics on upper layers, such as kbps at app layer, and app layer kbps is close to users' experiences.

MNOs might be lacking app layer measures, which are close to users' experiences. MNOs might do relationship study of lower layer throughput v.s. lower layer radio metrics (e.g., rsrp, rsrq, etc.). In some cases, if rsrp, rsrq are tuned to optimize lower layer's throughput, it is still unknown or unclear, whether the end users would be happy, how users experiences would be, whether users experiences could be predicted or estimated post tuning of the rsrp or rsrp, and whether users experiences would be enhanced. MNOs might want to find answers, but they might be limited to do so, as they might be lacking the upper layer data.

The ML-powered optimization in this article, solves such challenges. The approaches in this article can analyze and predict the tuning of the lower layers (e.g., rsrp, rsrq) directly to improve user's experiences, such as app layer kbps, user experienced latency, etc. ML models are established to find out how sensitive the tuning of rsrp, rsrq, would be for improving app layer kbps, or rtt user experiences. The models provide predicted improvement of app layer kbps, or rtt, etc., if the lower layer is tuned by a certain amount. The approaches directly translate lower layers tuning towards the end users' happiness (experiences).

The sensitivity analysis uses predictive modeling, to analyze the relationship between metrics. The models can find out local sensitivity, which can be indicted as a local slope, if linear regression is used, for a portion or in interval of the metrics X, v.s. the objective metrics Y.

One of the advantages is the slope, which may be used to indicate the sensitivity, could be used to identify the root causes, or the tuning knobs which are mostly sensitive to the objective function or objective metric Y. For an objective metric Y to be optimized (e.g. Y being app layer download speed), among other metrics X1, X2, X3, …, if X2 is most sensitive to Y (indicated by steepest slope in a relevant tuning interval of X2), while X1 or X3 does not have the high sensitivity towards Y, then the possible cause might be X2, or in other words, the highest priority to being the tuning knob is X2.

The sensitivities, e.g., the local slopes, can be used to prioritize which cells to be optimized first. Cells with higher sensitivity of lower layer metrics such as rsrp, rsrq, towards the upper layer metrics such as download speed, can be for higher priority to be optimized first.

As illustrated in Fig. 3, examples of the sensitivities of tuning the rsrp v.s. download speed, and rtt, respectively, are provided. In the example, the slopes can be observed, with steeper slopes as more sensitive, and less steeper slopes as not as sensitive. Cells with higher sensitivities can be optimized with higher priority of improving lower layer metrics such as rsrp, rsrq. In Fig. 3, the download speed is more sensitive to rsrq in cell_b than in cell_a, and the latency is more sensitive to rsrq in cell_d than in cell_c, especially in the regime of low to medium rsrq.

Another advantage of sensitivity analysis is the prediction would give out a predicted improvement of metric Y, if metric X is improved by a certain amount. This provides opportunities and advantages to do proactive optimization, with predicted upper layer end-users' experiences.

# Network optimization: RF parameters tuning; assisted by locations and directly targeting to improve users experiences

In wireless networks, coverage and capacity optimization is one of the most valuable problems to solve, and it is very difficult to solve it well. Telco industry has been in demand for better optimization solutions, which can lead to higher spectrum efficiency, hence more efficient network, and better experiences for users.

The coverage and capacity optimization face challenges. For example, for optimization solutions where the data of lower layer metrics such as rsrp, rsrq are used, if the data are not associated with geo-location such as latitude and longitude, the tuning might be more like blind of where end users are, and hard to tune. If the lower layer metrics are not associated with upper layer metrics, it might be hard to know if tuning rsrp, rsrq, what to expect for users experiences change, and sometimes it might happen even if the lower layer metrics are improved but the higher layer metrics are not improved much, due to low sensitivity of tuning of the lower layer metrics towards the higher layer metrics.

With the data collection as described as in Fig. 1, for a data set where the data include lower layers RF metrics such as rsrp (possibly indicating coverage), rsrq (possibly indicating capacity), etc., and higher layers performance and experiences metrics such as user experienced download speed, latency, etc., as well as the geo-location tags (geo location of latitude/longitude associated with each of the measures of these metrics), the data can be combined in a unique way to offer solutions and tools and potentially to achieve better network optimization and network performances, better users experiences on connectivity, and solving the challenges mentioned above.

With the power of such a data set as described in Fig. 1, sophisticated ML powered modeling and solutions can be built. The modeling include, e.g.,
1. Models which predict the improvement of upper layer metrics if tuning lower layer metrics, hence sensitivity is derived for upper layer metrics (e.g., download speed, latency) v.s. lower layer metrics (e.g., rsrp, rsrq).
2. Models for geolocation data based cell center, distance from the site to cell center, site to cell range, tuning recommendations for antenna azimuth based on the geo-distribution of the data samples logged, tuning recommendations for tilt and tx power based on the antenna heights, the distance of the site to cell range, the cell radius, and the geo-distribution of the data samples logged.
3. Models for iterative approaches for optimizing towards better distribution of rsrp, rsrq (e.g., reduce the tail probability), and for optimizing uses experiences at upper layer such as kbps, rtt, where at each step, a set of cells are chosen to be tuned based on the rsrp, rsrq, kbps, rtt, and the sensitivity of kbps, rtt v.s. rsrp, rsrq. Most sensitive ones, worst rsrp, rsrq, kbps, rtt, would have higher priorities. Local cells of the tuned cells are jointly optimized, based on the cell radius, site to cell range and the geo-distribution of the data samples.

Figure 4 illustrates an example of network optimization with tuning of RF parameters, where antenna azimuth tuning is used as an example.

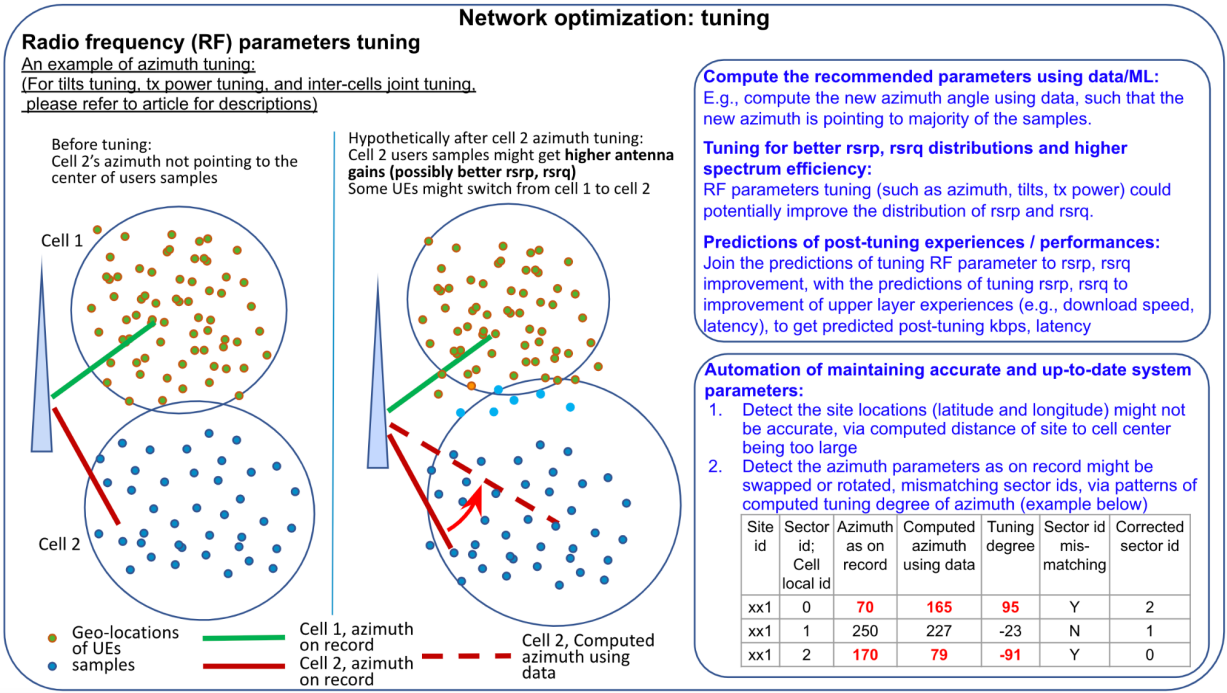

*Fig. 4 An example of network optimization with tuning of RF parameters, where antenna azimuth tuning is used as an example.*

In Fig. 4, there is an example of illustration of the visualization of antenna azimuth tuning case, with before tuning and after tuning comparison. In the figure, the dots represent the geo-locations of end-users data samples. Before tuning, Cell 2's azimuth is not pointing to the center of the data samples. Hypothetically after Cell 2 azimuth tuning, Cell 2 data samples might get higher antenna gains (possibly better RSRP, RSRQ). Some UEs at the edge of both cells might switch from Cell 1 to Cell 2 as Cell_2 might provide better RSRP and RSRQ post tuning of Cell_2's azimuth.

Data analytics and machine learning can be used to compute the recommended parameters. For example, a new and recommended azimuth angle is computed using data, such that the new azimuth is pointing to the majority of the samples. A center of the data samples with geo locations can be first computed, with the data samples associated with a certain cell. Then, the new azimuth angle can be the angle pointing to the center of the data samples.

If the current azimuth angle is known, then, a further calculation can be made, to determine the tuning angle of the azimuth, where the tuning angle can be the difference between the current azimuth to the new and recommended azimuth.

Using machine learning method, the predictions of post-tuning network performances and user experiences can be done by joining the predictions of tuning RF parameter to rsrp, rsrq improvement, and the predictions of tuning rsrp, rsrq to improvement of upper layer experiences

(e.g., download speed, latency), to get the predicted post-tuning kbps, latency. For example, in the first step, find out rsrp gains, due to increased antenna gain, per a certain tuning degree of the antenna azimuth [5]. Then, in the second step, by looking into the regression model's results, such as a local slope of the sensitivity of rsrp v.s. upper layer metrics like download speed, latency, as the plot in Fig. 3, the predicted upper layer metrics improvement can be estimated.

In addition, the automation of maintaining accurate and up-to-date system parameters may be possibly done, in some cases.

In one case, for example, using the data and ML models, it may be detected that the site locations (latitude and longitude) might not be accurate (e.g., might be a manual mistake by engineers who record the site locations parameters to a spreadsheet of the system parameters), via computed distance of site to cell center being too large. In this case, a warning or an alert for double checking the site locations parameters can be generated. The event of the computed distance of site location to cell center location being too large, can be indicated by the distance being larger than a certain threshold, and possibly jointly with other conditions. On a site, if there are multiple cells (e.g., 3 cells), and all these cells have this event detected, then it could be an even stronger indication that the site location parameters are inaccurate, than the case where only one cell of the site has such an event detected.

In another case, for example, using the data and ML models, it may be detected that the azimuth parameters as on record might be swapped or rotated, mismatching sector identifiers, via patterns of computed tuning degree of azimuth, and the relationship of the estimated azimuth tuning degree v.s. the sector identifier or cell local identifier.

In Fig. 4, in the example provided, for the same site identifier xx1, there are three sectors, or three cells. Cell_0 and Cell_2 are recommended to be tuning the azimuth 95 degree clockwise, and 91 degree counterclockwise, respectively. As these two tuning degrees are of similar amount of degree and with opposite directions, it indicates the sector id is mismatching the parameter configuration of the azimuth, and indicating this might be a situation that the actual hardware on these two sectors might have been swapped, e.g., during the phase when the engineers installed or deployed the sectors. If the hardware pieces are similar for sector id being 0 and 2, the engineers might not need to reinstall the hardware (swapping them), rather, they might need to correct the parameters sheet, i.e., in this case, the hardware with sector id being 0 should be correctly recorded as pointing to the azimuth direction of 170 degree.

The optimization solutions and tools can be built, for example, via "a browser based network optimizer". Optimization recommendations can be computed. On the browser, customers can create/save optimization projects, select the cells in a local area to be optimized, visually see the geo-distribution of traffic demand, and determine the actions (tuning). The customers could record the actual actions on the browser. The post-tuning performance can be monitored via the browser (e.g., daily, via time series), recommendations can be updated to the browser after recomputing, and the parameters might be tuned again if needed, hence this type of tuning can be iterative, until the performances are optimized for the selected area. The automation of maintaining accurate and up-to-date system parameters mentioned above may be also facilitated via the browser.

# ML-Assisted Wireless Network Planning

Network planning could have two phases in general, one is the initial planning, and the other is network expansion/upgrading after the deployment.

Initial network planning can refer to the planning for the wireless networks in a certain area, for the initial deployment. It can include radio technologies selections, backhaul technologies selections, site selections, site density analysis, service provisioning analysis, end-to-end quality of experiences provisioning, etc.

Network densification/upgrading can refer to the planning after the initial deployment within an area. It can include, e.g., detection the areas or opportunities for adding more cells, upgrading the cells such as upgrading to more advanced technologies, more advanced hardware, newer generation of the technologies, etc., and then predict the performances if adding or upgrading, and recommend the plan for the actual deployment of the addition, upgrading, etc. Ideally, network optimization based on software or more soft solutions like tuning etc. may be done first, before the actual network densification/upgrading.

When telecom is evolving generation after generation, now more and more 5G, even considering the future 6G, with the potential future Open RAN (ORAN) architecture, and with emerging use cases (e.g. Augmented Reality (AR), Autonomous Cars, and other next generation use cases ), a fundamental question is what we can do differently to disrupt the traditional network planning with ML/AI, to utilize the strength of the data, as well as ML/AI tools, to help boost the advantages in terms of e.g., the performance gains such as in end-users experiences,network efficiency, network reliability, and computational cost saving from the economic perspective.

The ML powered network planning modules may include demand modeling, the understanding of the requirements on QoS/QoE, the modeling of the current capacity the network can provide, and detect whether the current capacity of the network is no longer sufficient to the demand, the predictive modeling on determining where/how to upgrade to new hardware /technology, or add new nodes, with predicted performances, and so on. Subsections below provide some of more details, also with consideration of the future ORAN perspectives.

## ML-powered demand modeling, requirements understanding

The demand modeling is among the most important modules for ML-powered network planning. It is important to both initial planning and network densification/upgrading.

As the data collection as described in earlier sections, the deidentified location data can be tagged to the performance and experiences data, as well as the total traffic bytes and number of data samples. With such, demand modeling can be performed, utilizing the data set, from temporal and spatial perspectives.

The data-driven demand modeling can provide the understanding of the current demand, and future demand trends which can be built via predictions with the historical data, also possibly with the other data sources regarding the future applications, etc. The demand modeling can also provide the spatial distribution of the demand, which could be utilized together with the maps, to provide better understanding of prioritization of areas, hot spots, and site locations, etc.

Beyond temporal and spatial perspective for demand modeling, another important perspective is about the various applications the area is serving, traffic volume of each application in terms of the proportion of the total traffic, and for the applications, what kind of requirements of QoS/QoE for each application. The requirements can include, e.g., the requirements of application level of the download speed, user's end-to-end round trip time, etc.

For example, in some areas, the live video traffic might take L% or above in the total amount of traffic, where the live video should better have a median latency <= T ms, and then for these areas, the network should be planned to meet such requirements. Similarly, for augmented reality (AR) services, gaming services, messenger live video chat services, etc., the requirements should be also understood, together with the proportion of such applications traffic volume and priorities, or importance from the marketing perspectives.

Hence, the demand modeling, in its nature, is data driven, and ML-powered, with predictive modeling, spatial and temporal modeling, as well as detailed per application requirements analysis and predictions.

## ML-powered network planning with predictive metrics

The ML-powered network planning modules may also include the modeling of the current capacity the network can provide, and detect whether the current capacity of the network is no longer sufficient to the demand, hence the end-users demand has been suppressed. The planning modules may also include the predictive modeling on determining where/how to upgrade to new hardware /technology, or add new nodes, with predicted performances.

One of the most critical problems for network densification/upgrading, is to detect the need for network densification/upgrading and to justify why network densification/upgrading is needed, what is the predicted performances or end-users' experiences if densification or upgrading is done, with how much incentive or return of investment (ROI) if the network densification/upgrading is done.

Approaches to solve this critical problem of why network densification/upgrading is necessary include the engagement suppression analysis, as it can particularly predict and quantify how much engagement or how much traffic is suppressed.

Engagement suppression is to identify and predict how much end-users' engagement is suppressed, due to e.g., bad experiences (e.g., bad video watching experiences due to bad connections, congestions), etc. The suppressed engagement could be the suppressed engagement time duration, or it could be the suppressed traffic, which is the traffic which could have been consumed by the end user but actually it is not consumed, due to e.g., bad video playback experiences due to bad connection, congestions, etc.), etc. For instance, some end user may purposely decide not to watch high definition video they are interested, because of the bad quality of the wireless connection, but if the connection were good, the end user would have watched it.

Figure 5 shows an illustration of deriving predictive metrics for network planning purposes, from some perspectives of densification and upgrading, where the predictions with consideration of recovering of the suppressed demand is also illustrated.

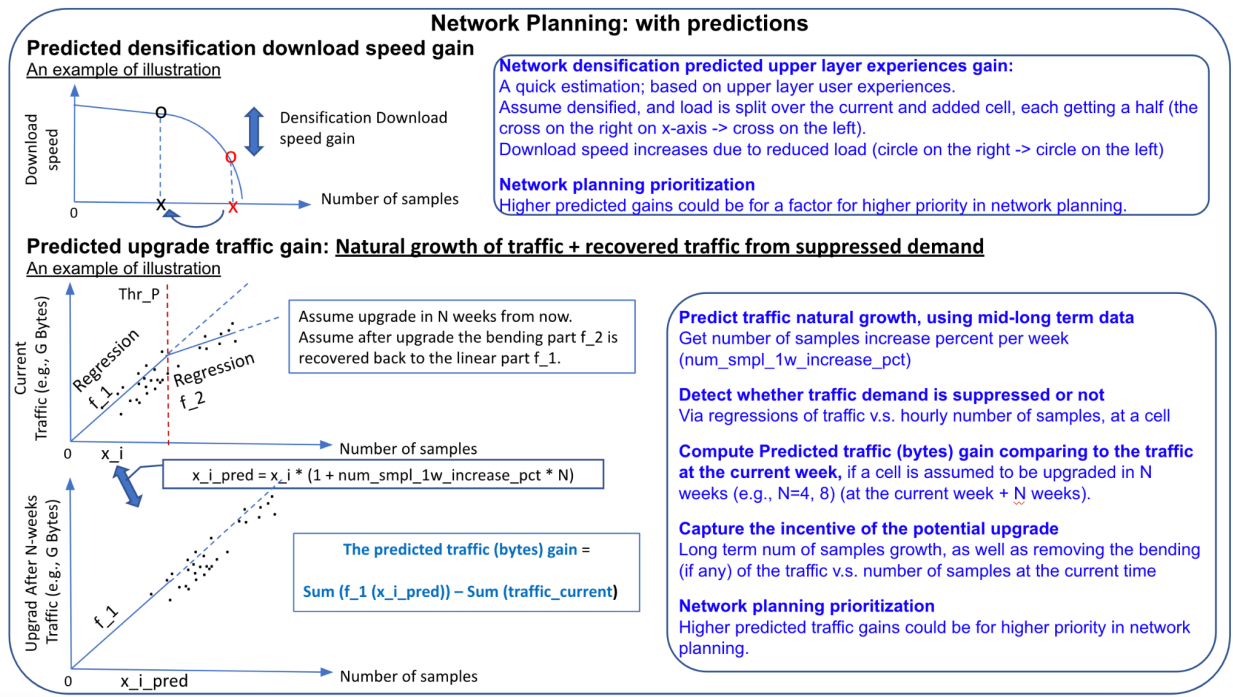

*Fig. 5 An illustration of deriving predictive metrics for network planning purposes, from some perspectives of densification and upgrading, where the predictions with consideration of recovering of the suppressed demand is also illustrated.*

In Fig. 5, it shows an illustration of the network densification predicted upper layer experiences gain. This gain is a quick estimation; based on upper layer user experiences. The download speed case is shown as the example. Assume the cell is densified, and load is split over the current cell and the added cell, each getting a half (for the current cell, the load is reduced from the cross on the right on x-axis to the cross on the left). Then, the app layer download speed increases due to reduced load (for the current cell, the download speed increases from the circle on the right to the circle on the left).

This metric of predicted densification download speed gain, can be used in network planning prioritization, where the cells or areas with higher predicted gains could be a factor for higher priority in network planning. Similarly, other predicted metrics of the gains due to densification can be derived, and utilized to identify or detect the cells or areas which can be great candidates for densification.

In Fig. 5, it also shows an illustration of the predicted upgrade traffic gain, which takes into account both the natural growth of traffic and the recovered traffic from suppressed demand.

The predicted traffic natural growth is from ML models using mid-long term data. From the ML models, the number of samples' increase percent per week (num_smpl_1w_increase_pct) can be derived.

ML models can be built to detect whether traffic demand is suppressed or not. For example, the detection can be done via the regressions of traffic v.s. hourly number of samples, at a cell. In Fig. 5, each data point represents the hourly data, with x-axis being the total number of samples within an hour, and y-axis being the total bytes of the traffic of these samples within this respective hour. A first regression is made, with the hourly data points with lower numbers of samples. A second regression is also made, for the hourly data points with higher number of samples. The second regression has a flatter slope than the first regression, meaning that the total traffic demand is bending, w.r.t the number of total samples contributing to the total traffic. In other words, it could be possible that the end-users are more likely clicking or requesting the content with smaller response body size, when the network is busy, indicating that the demand at the end users might have been suppressed.

If the similar behavior of requesting content at the non-busy hours persists towards at the busy hours, then the second regression should have been with the similar slope as the first regression, not bending noticeably or even significantly.

Then, predicted upgrade traffic (bytes) gain can be computed, compared to the traffic at the current week, if a cell is assumed to be upgraded in N weeks (e.g., N=4, 8) (at the current week + N weeks). Due to the natural growth, in N weeks, the number of samples at the x-axis would be shifted towards right, as $x\_i\_pred = x\_i * (1 + num\_smpl\_1w\_increase\_pct * N)$. Then, the The predicted upgrade traffic (bytes) gain = Sum (f_1 (x_i_pred)) – Sum (traffic_current).

This metric captures the long term growth of the number of samples, as well as removing the bending (if any) of the traffic v.s. number of samples at the current time.

This metric of predicted upgrade traffic gain may reflect the incentive of the potential upgrade, in terms of the gained traffic in bytes. If the traffic in bytes could be related to how the MNO is charging the end user, then this metric could be related to MNO's revenue incentive of the potential upgrade.

This metric of predicted upgrade traffic gain can be used in network planning prioritization. Higher predicted traffic gains could be for higher priority in network planning of densification as well as upgrading.

Another metric, not included in Fig. 5, but the concept could be understood via Fig. 5, is the time for a cell to run out of capacity. It is an inverse of the predicted upgrade traffic gain in terms of percent. If the predicted upgrade traffic gain in percentage is hitting beyond a certain threshold, e.g., 5%, then the time can be derived for such hitting. The time could be, e.g., in how many weeks. If at the current week, the gain is already beyond 5%, then the week to run out of capacity is zero. The predicted time for a cell to run out of capacity can be used to indicate how urgent an upgrade of a cell's capacity is needed.

## Open RAN Insights (ORANi)

Open RAN (ORAN) is a new direction of the RAN (radio access network) architecture in telecom industry [6, 7]. It opens up certain interfaces for more flexibility, for lower TCO (total cost of ownership), better management, better virtualization, better resource sharing, better intelligence in RAN, etc. The serviceable available market share of ORAN seems to have great potential with all ORAN solutions.

ORAN Insights (ORANi) is to use data collected from app API, as well as other data, and data analytics, machine learning and artificial intelligence, with wireless system domain knowledge and expertise, to help provide solutions to ORAN.

With ORANi, traffic demand modeling and traffic geographic distribution modeling can be used to identify where the opportunities for ORAN solutions are. Given a certain amount of budget (budget limit), certain areas are of higher priority for ORAN solutions, such as predicted high demand areas, the areas worthy of ORAN investment.

With ORANi, predictive modeling can be used to detect and predict where there are the opportunities for massive MIMO. Cells with massive MIMO units could be of higher computational cost from a hardware perspective. Given a certain amount of budget (budget limit), certain cells or sites may be of higher priorities to have massive MIMO capability (e.g., predicted higher traffic demand of cells). Certain areas may be of higher priorities to have the hardware of massive MIMO. The prioritization of an area whether massive MIMO units should be deployed, can be determined by, for example, using the predicted upgrade gain, as in FIG. 5. The higher predicted upgrade gain, the higher priority for upgrading with massive MIMO, especially if the predicted upgrade gain is in the unit of GigaBytes (GB).

# Open Problems and Future Research Directions

This section provides some examples of the open problems and future research directions.

One of them is regarding how to jointly utilize the strength of data from different parties. Some parties might have some data which have strength from various perspectives, such as different OSI layers, and different focuses. Sometimes if the strength of the various data could be utilized jointly, the solutions might be more powerful and more effective.

Another of them is regarding how to utilize the data, and build up models and solutions, to do the cross layer optimization in a better way, such that different layers information could be better utilized, especially considering the different time scale of the optimization, as well as the time consuming by data collection, processing, etc.

Another of them is regarding how to utilize various data, e.g., the measurements of the radio signal strength and quality, the maps information, 2D/3D digital maps, etc., to come up with the in-depth modeling regarding the wireless environment, as well as the predictions of signal strength, signal quality, propagation, etc.

Yet another of them is how to jointly utilize various data, including maps data, to come up with the models and solutions, to automatically figure out or recommend where exactly to mount the radio access point, without engineers much involvement.

# Conclusion

To serve the explosively growing traffic and diverse applications requirements of throughput, latency, and reliability, it is critical that the wireless network is well managed, optimized and planned, using the limited wireless spectrum resources. Considering the challenges of dynamics and complexity of the wireless systems, and the scale of the networks, it is important to have solutions to automatically monitor, analyze, optimize, and plan the network, instead of the traditional way of engineers manually monitoring, analyzing, tuning, and planning the network.

This article discusses approaches and solutions of data analytics and machine learning (ML) powered optimization and planning, to address the challenges of existing network optimization and planning technologies.

The approaches discussed include analyzing some important metrics of performances and experiences, at the lower layers and upper layers of open systems interconnection (OSI) model. The approaches include deriving a metric of the end user perceived network congestion indicator. The approaches also include monitoring and diagnosis such as anomaly detection of the metrics, root cause analysis for poor performances and experiences.

The approaches also include enabling network optimization with tuning recommendations, directly targeting to optimize the end users experiences, via sensitivity modeling and analysis of the upper layer metrics of the end users experiences v.s. the improvement of the lower layers metrics due to tuning the hardware configurations.

The approaches also include deriving predictive metrics for network planning, and modeling of traffic demand distributions and trends, incentives of traffic gains if the network is upgraded, etc. The models detect and predict the suppressed engagement or suppressed traffic demand.

These approaches of optimization and planning may provide more accurate detection of optimization and upgrading/planning opportunities for cells at a large scale, enable more effective optimization/planning of networks, such as tuning cells configurations, upgrading cells' capacity with more advanced technologies or new hardware, adding more cells, etc., improving the network performances and providing better experiences to end users of the networks.

Future work and directions include how to jointly utilize the strength of data from different parties, how to utilize the data, and build up models and solutions, to do the cross layer optimization in a better way, such that different layers information could be better utilized, how to utilize various data, e.g., the measurements of the radio signal strength and quality, the maps information, 2D/3D digital maps, etc., to come up with the in-depth modeling regarding the wireless environment, as well as the predictions of signal strength, signal quality, etc., and to come up with the models and solutions, to automatically figure out or recommend where exactly to mount the radio access point, without engineers much involvement.

# Acknowledgements

The authors cordially appreciate all the discussions with their colleagues.

# Authors

**Dr. Ying Li** received the M.A. and Ph.D. degrees in Electrical Engineering at Princeton University, Princeton, NJ, in 2005 and 2008, respectively.

Her recent research focuses on big data, applied machine learning and artificial intelligence, and their applications on connectivity, network optimization and planning, ads targeting, spatial computing, etc. She has been with Meta Platform since July 2017. Prior to that, she worked with FutureWei Technologies (2015 - 2017), Bridgewater, NJ, where she worked on data analytics and machine learning for wireless networks solutions. She was with Samsung Research America (2013 - 2015) and Samsung Telecommunications America (2008 - 2013), Dallas, TX, where she conducted research on communication networks and smart energy networks. She was a visiting scholar in Swiss Federal Institute of Technology (EPFL), Lausanne, Switzerland, in summer 2007 and in Motorola Multimedia Research Labs, Schaumburg, IL, in fall 2007, respectively.

Dr. Li is an inventor of 75 patents granted by USPTO as of Dec. 2021 (including 4G/5G standards/products patents). She is a recipient of Future Star Award from FutureWei Technologies in 2016, Distinguished Inventor Award 2013 from Samsung Research America, and the Inventor of the Year 2012 and Distinguished Inventor Award 2010 from Samsung Telecommunications America. She is a recipient of Gordon Wu Fellowship from Princeton University in 2003-2007, for new graduate students who demonstrate the potential to be world leaders in their fields in the 21st century.

**Dr. Djordje Tujkovic** received the Ph.D. degree in wireless communications from the University of Oulu, Finland, in 2003. From 2003 to 2004, he was a Post-Doctoral Scholar and the Research Project Manager with the Centre for Wireless Communications, University of Oulu. From 2004 to 2005, he was a Post-Doctoral Fellow with the Smart Antenna Research Group of Prof. Arogyaswami Paulraj, Standford University. In 2004, he joined Beceem Communications, Santa Clara, CA, USA, an early stage WiMax and Cellular silicon startup, where he was the Director of engineering until acquisition of Beceem by Broadcom in 2010. He continued to work on wireless communication system and chip design at Broadcom as an Associate Technical Director. In 2014, he joined the Connectivity Lab, Facebook Inc., where he is currently the Director of engineering and a System Lead on mmWave wireless distribution network program.

**Dr. Po-Han Huang** received the M.S. degree in computer science and Ph.D. degree in computer engineering from the University of Southern California (USC), Los Angeles, CA, in 2018 and 2020, respectively. His research interests include algorithm design, performance modeling, optimization, and planning in wireless networking, datacenter networking, and vehicular networking, and applied machine learning and artificial intelligence on networking/connectivity. He has been with Meta Connectivity since May 2020. He was with Nokia Bell Labs, Murray Hill, NJ and with Intel Labs, Santa Clara, CA in summer 2017 and 2018, respectively. He is a recipient of the Annenberg Fellowship from USC in 2015-2019 and the Jenny Wang Excellence in Teaching Award from USC Viterbi School of Engineering in 2019.